\documentclass{article}
\usepackage{graphicx}

\begin{document}
\title{Non-relativistic Geodesic Behaviors for a Massive Charged Particle Falling in de Sitter Spacetime}

\author{Farrin Payandeh$^1$\thanks{e-mail:
f$\_$payandeh@pnu.ac.ir} ,
 Mohsen Fathi$^2$\thanks{e-mail: mohsen.fathi@gmail.com}}
\maketitle \centerline{\it $^{1}$Department of Physics, Payame
Noor University, PO BOX 19395-3697, Tehran, Iran}

\centerline{\it$^2$Department of Physics, Islamic Azad University,
Central Tehran Branch, Tehran, Iran}

\begin{abstract}

In this article, continuing the work done in the previous paper
(M. Fathi 2012), we apply a Lagrangian formalism to demonstrate
the shape of the geodesic motion for a massive charged particle
which is falling freely in a de Sitter spacetime. We will show
that a spiral shape of the trajectory is available, due to the
logarithmic behavior of time, with respect to the proper time.

\end{abstract}

\section{Introduction}
As well as an accelerating charge can radiate electromagnetic
energy, possessing a momentum, this emission can affect the
particle's trajectory because of a side effect, called radiation
reaction. The force, caused by a rate of change in this momentum,
is known as Abraham-Lorentz force \cite{abraham}. The radiation
reaction is a recoiling force and is proportional to the rate of
change in acceleration. Afterwards, Dirac employed space-like
geodesics to the Abraham-Lorentz force to generalize it to
relativistic velocities. This force is called the
Abraham-Lorentz-Dirac force \cite{poisson}.

In this work, connected to the derivations in \cite{Fathi}, we use
a Lagrangian formalism, for a massive charged particle, to derive
the relations between the coordinate time and the affine parameter
of the trajectory. This affine parameter, will be considered as
the proper time. Like before, we choose the flat coordinate system
in a de Sitter space time and the effective potential will be
initially derived. The test particle is considered to be finitely
small, to avoid impositions of the side effects of irregularity.

Therefore, the relation for the coordinate time with respect to
the proper time, will be derived form the Lagrange equations. Then
using numerical simulations, the shape of the possible orbit will
be illustrated, showing how the particle approximately behaves
while it is falling freely in a de Sitter spacetime.

The paper is organized as follows: In section two, bring a brief
review of the flat coordinate system in de Sitter spacetime and
the effective potential for a freely falling particle in this
spacetime, as well as the geodesic equations, are derived. In
section three, we give the total external force (including the
radiation reaction), causing the total kinetic energy of the test
particle. In section four, the Lagrangian formalism is presented
and the numerical simulations are given.

\section{de Sitter Spacetime}
The de Sitter spacetime is a vacuum solution for Einstein
equations with a positive cosmological constant $\Lambda$,
\begin{equation}
 R_{\mu\nu}- \frac{1}{2} R g_{\mu\nu} + \Lambda g_{\mu\nu} = 0.
 \label{Enstein equation}
 \end{equation}
The de Sitter spacetime is the unique maximally symmetric curved
spacetime characterized by the following condition \cite{Winberg,
takook}:
\begin{equation}
R_{\mu\nu\lambda\rho} = \frac{\Lambda}{3}
(g_{\mu\lambda}g_{\nu\rho}- g_{\mu\rho}g_{\nu\lambda}),
 \label{Riemann tensor}
 \end{equation}
where $R_{\mu\nu\lambda\rho}$ is the Riemann curvature tensor. The
relations $R_{\mu\nu} = R^{\lambda}_{\mu\lambda\nu}$ , $R =
g_{\mu\nu} R^{\mu\nu}$ lead to:
\begin{equation}
R = 4 \Lambda,
 \label{Ricci scalar}
 \end{equation}
in which $R$ is the Ricci scalar. The metric in de Sitter
spacetime is defined as follows:
\begin{equation}
ds^2 = g^{ds}_{\mu\nu} dX^\mu dX^\nu, \quad\mu , \nu = 0,1,2,3
 \label{de sitter metric general}
 \end{equation}
in which $X^\mu$ is the four-dimensional intrinsic coordinates of
this four dimensional hyperbolic spacetime. The de Sitter
spacetime is a hyperboloid space, embedded in a five dimensional
space, called the Ambient space:
\begin{equation}
X_H = \{x \in R^5 ; x^2 = \eta_{\alpha\beta}x^\alpha x^\beta = -
H^{-2}\},\quad
 \alpha , \beta = 0,1,2,3,4
  \label{X_H}
 \end{equation}
for which the metric is given by:
\begin{equation}
ds^2 = \eta_{\alpha\beta} dx^\alpha dx^\beta, \quad
 \eta_{\alpha\beta} = diag\{1,-1,-1,-1,-1\}
\label{de Sitter metric - 5D}
\end{equation}
Here $H^{- 1}$ is the minimum radius of the hyperboloid and $H$ is
the Hubble parameter. Note that, in this paper, the Hubble
parameter will be regarded as a constant (Hubble constant), but
some analytical solutions for the Hubble parameter and the
cosmological constant, have been presented within other
gravitational theories (see \cite{tanhayi, Fathi1}). The most
important coordinate systems in de Sitter spactime are: Global,
Conformal, Flat and Static \cite{takook}. We use the flat
coordinate system, which is defined as follows:
\begin{equation}
ds^2 = c^2dt^2 - e^{2Ht}dX_i^2,\,\,\,\,\,\,i=1,2,3.
 \label{de Sitter flat metric}
 \end{equation}
We use this metric for further calculations.

\subsection{the effective potential}

For a test particle, having charge $q$ and mass $m$, the equations
of motion can be derived from the Hamilton-Jacobi equations of
wave crests \cite{Wheeler}:
\begin{equation}
g^{\mu\nu}P_\mu P_\nu+m^2=0,
 \label{HG-1}
 \end{equation}
in which $\lambda$ is the affine parameter and the 4-momentum
$P_\mu$ is defined as:
$$P_\mu=g_{\mu\nu}P^\nu=g_{\mu\nu}\frac{dX_{\nu}}{d\lambda}.$$
Here, the charge $q$, will temporarily disappear, since the
particle is falling freely. However, in our further
considerations, where the radiation reaction comes into account,
the charge $q$ receives an important roll in our calculations.
Using the metric (\ref{de Sitter flat metric}) in (\ref{HG-1}) and
defining the energy as $E=P_0$, we get:
\begin{equation}
E^2-e^{2Ht}\sum_{i=1}^3(\frac{dX_i}{d\lambda})^2+m^2=0.
 \label{HG-2}
 \end{equation}
If we take the spatial coordinates $X_i$, indistinguishable, then
we deduce form (\ref{HG-2}) that:
\begin{equation}
E^2-3e^{2Ht}(\frac{dX}{d\lambda})^2+m^2=0,
 \label{HG-3}
 \end{equation}
or
\begin{equation}
(\frac{dX}{d\lambda})^2=\frac{E^2+m^2}{3e^{2Ht}},
 \label{HG-3}
 \end{equation}
from which we introduce the effective potential as:
\begin{equation}
V_{eff}=\frac{m}{\sqrt{3}}e^{-Ht}.
 \label{Veff}
 \end{equation}
This is the effective potential which the test particle feels
during its geodesic motion. Figure 1 shows the behavior of this
time-dependent potential.
\begin{figure}[htp]
\center{\includegraphics[width=6cm]{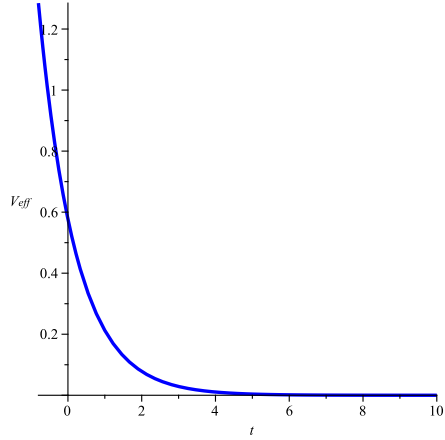} \caption{\small{The
time-dependent potential for a freely falling particle in de
Sitter space-time}}} \label{EP}
\end{figure}

\subsection{the geodesics}

To derive a relation between the coordinate time and the spatial
coordinates in (\ref{de Sitter flat metric}), we use the geodesic
equations:
\begin{equation}
\frac{d^2X^{\mu}}{d\lambda^2} + \Gamma^{\mu}_{\nu\rho}
\frac{dX^\nu}{d\lambda} \frac{dX^\rho}{d\lambda} = 0,
 \label{Geodesic equations formula}
 \end{equation}
which leads to:
\begin{equation}
\frac{d}{d\lambda}X(\lambda)=e^{-2Ht(\lambda)},
 \label{e1}
 \end{equation}
\begin{equation}
\frac{d^2}{d\lambda^2}t(\lambda)=-\frac{3}{c^2}He^{-2Ht(\lambda)}.
\label{e2}
 \end{equation}
The test particle's velocity can be derived using (\ref{e1}) and
(\ref{e2}) as follows \cite{Fathi}:
\begin{equation}
v=\frac{ce^{-Ht}}{\sqrt{3+e^{2Ht}\Big(-3+\frac{c^2}{v_0^2}\Big)}}.
\label{v}
\end{equation}
It is easy to show that the following constraint is imposed on the
initial velocity $v_0$:
$$v_0<\frac{c}{\sqrt{3(1-e^{-2Ht})}}.$$
Therefore we can obtain the relation between the coordinate time
and the spatial coordinates as:
\begin{equation}
X(t)=\frac{c}{3H}\Big(\frac{c}{v_0}-e^{-Ht}\sqrt{3+(-3+\frac{c^2}{v_0^2})e^{2Ht}}\,\,\Big).
\label{tr}
\end{equation}
In the next section we introduce the radiation reaction and use it
to illustrate the effects of the resultant force, on particle's
trajectory.

\section{Radiation Reaction}
Now we present a brief review on the radiation reaction. The
radiation reaction provides a recoiling force which is given by
\cite{f}:
\begin{equation}
F_{re}^\mu=\frac{2}{3}\frac{q^2}{c^3}\ddot v^\mu+
\frac{2}{3}\frac{q^2}{c^3}v^\mu \dot v^\lambda \dot v_\lambda,
\label{j1}
\end{equation}
where $v$ represents the test particle's velocity. The
Lorentz-Dirac equation which relates the particle's motion and the
external force is given by:
\begin{equation} \label{ld}
ma^\mu= F^\mu_{ext}+F^\mu_{re},
\end{equation}
in which $a^\mu$ is the 4-acceleration. Considering
$\frac{q^2}{mc^3}$ to be a small valued quantity and using
(\ref{j1}) one obtains \cite{poisson, f2, landa}:
\begin{equation}
ma^\mu=F_{ext}^\mu+\frac{2}{3}\frac{q^2}{c^3m}\Big(\delta_\lambda^\mu+v^\mu
v_\lambda\Big)\partial_\gamma F^\lambda_{ext}v^\gamma, \label{2d}
\end{equation}
In non-relativistic limit (\ref{2d}) turns to \cite{poisson,
landa}:
\begin{equation} \label{nr}
ma=F_{ext}+\frac{2}{3}\frac{q^2}{mc^3}\dot F_{ext}.
\end{equation}
The reader can find such relations in curved space in \cite{cu,
curved-po}.

We can rewrite (\ref{nr}) as $F_{total}\equiv ma$ for which we
consider $F_{re}=\frac{2}{3}\frac{q^2}{c^3}\ddot v$. For the
radiation reaction recoiling force, one can write \cite{Fathi}:
\begin{equation}
F_{re}=\sum^{\infty}_{n=1}\alpha^n(\frac{d^n}{dt^n})F_{ext},
\label{E15}
\end{equation}
in which $\alpha \equiv \frac{2}{3}\frac{ q^2}{mc^3}.$ for small
values for $\frac{q^2}{mc^3}$ we get:
\begin{equation}
F_{total}=\sum^{\infty}_{n=0}\alpha^n(\frac{d^n}{dt^n})F_{ext}.
\label{E16}
\end{equation}
In non-relativistic limits, where $v_0\ll c$, equation (\ref{tr})
leads to:
\begin{equation}
X(t)=\frac{v_0}{2H}\Big(1-e^{-2Ht}\Big). \label{trajectory}
\end{equation}
Therefore, the external force in this limit, turns to:
\begin{equation}
F_{ext}=m \ddot{X}(t)=-2mHv_0e^{-2Ht} \label{F-external}.
\end{equation}
Using (\ref{F-external}) in (\ref{E16}) we obtain \cite{Fathi}:
\begin{equation}
X(t)=\frac{v_0}{(2H)(1+2H\alpha)}\Big(1-e^{-2Ht}\Big).
\label{final-trajectory}
\end{equation}
This is the general equation which relates $X$ to coordinate time
$t$.

\section{Lagrangian Formalism}
The external force in (\ref{F-external}), provides an external
momentum which is being imposed on the test particle:
\begin{equation}
P_{ext}=\int F_{ext}dt=mv_0e^{-2Ht}. \label{P-external}
\end{equation}
This momentum is a source for the kinetic energy of the test
particle. We have:
\begin{equation}
K_{ext}=\frac{P^2_{ext}}{2m}=\frac{1}{2}mv_0^2e^{-4Ht}.
\label{P-external}
\end{equation}
Now let us construct a Lagrangian like:
\begin{equation}
\mathcal{L}=K_{ext}-V_{eff}, \label{Lagrangian1}
\end{equation}
in which $V_{eff}$ has been defined in (\ref{Veff}). In de Sitter
flat metric (\ref{de Sitter flat metric}), this Lagrangian is a
function like:
\begin{equation}
\mathcal{L}\equiv\mathcal{L}(t,X_1,X_2,X_3,\dot t, \dot X_1, \dot
X_2, \dot X_3). \label{Lagrangian3}
\end{equation}
Here the dot stands for differentiation with respect to affine
parameter $\lambda$ in geodesic motion. Using the metric (\ref{de
Sitter flat metric}) with $c=1$ yields:
\begin{equation}
\mathcal{L}=\frac{1}{2}\,m{{v_0}}^{2}{{e}^{-4\,H{t}}}-\frac{1}{3}\,m\sqrt
{3}{ {e}^{-H{t}}} . \label{Lagrangian4}
\end{equation}
Since we take the geometrical units ($c=1$), we have
$d\lambda=d\tau$, where $\tau$ is the proper time. For
indistinguishable spatial coordinates, the action in this space
time, is defined by \cite{Greiner}:
\begin{equation}
S=\int\mathcal{L}(t,X,\dot t,\dot X)d\tau, \label{action1}
\end{equation}
Varying this action, we can obtain the Euler-Lagrange equation of
motion in de Sitter spacetime:
\begin{equation}
\frac{\partial\mathcal{L}}{\partial
X^i}-\frac{d}{d\tau}\Big(\frac{\partial\mathcal{L}}{\partial\dot
X^i}\Big)=0. \label{E-L1}
\end{equation}
Since the metric contains only time-dependent expressions,
therefore only one equation will come out from (\ref{E-L1}):
\begin{equation}
6\,{{v_0}}^{2}{{e}^{-4\,Ht \left( \tau \right) }}-\sqrt {3}{{
e}^{-Ht \left( \tau \right) }}=0.
 \label{E-L1-Eq}
\end{equation}
As we can see, the mass $m$ has no contribution in time-dependent
equation. This leads to derive the following expression for $t$:
\begin{equation}
t(\tau)=\frac{\Big(2\ln(2v_0^2)+\ln(3)\Big)}{6H}
 \label{t-Eq}
\end{equation}
\subsection{shapes of the trajectory}
Now we can illustrate the coordinate behaviors. First, let us
concern about the coordinate time $t$, which has been shown that
it has a logarithmic behavior with respect to initial velocity
$v_0$. Using equation (\ref{t-Eq}), we can plot the test particles
trajectory, which is shown in Figure 2.
\begin{figure}[htp]
\center{\includegraphics[width=6cm]{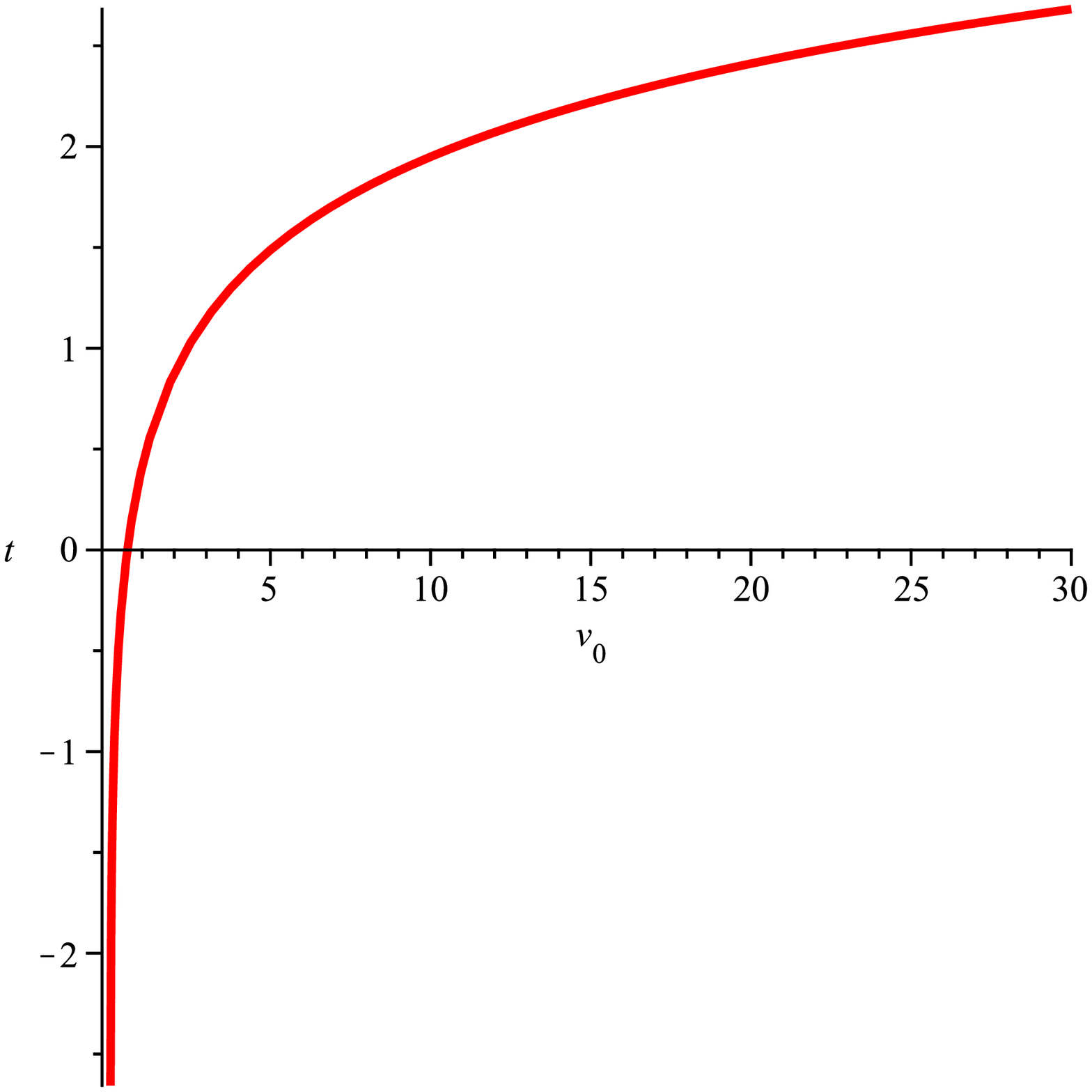} a)
\hfil
\includegraphics[width=6cm]{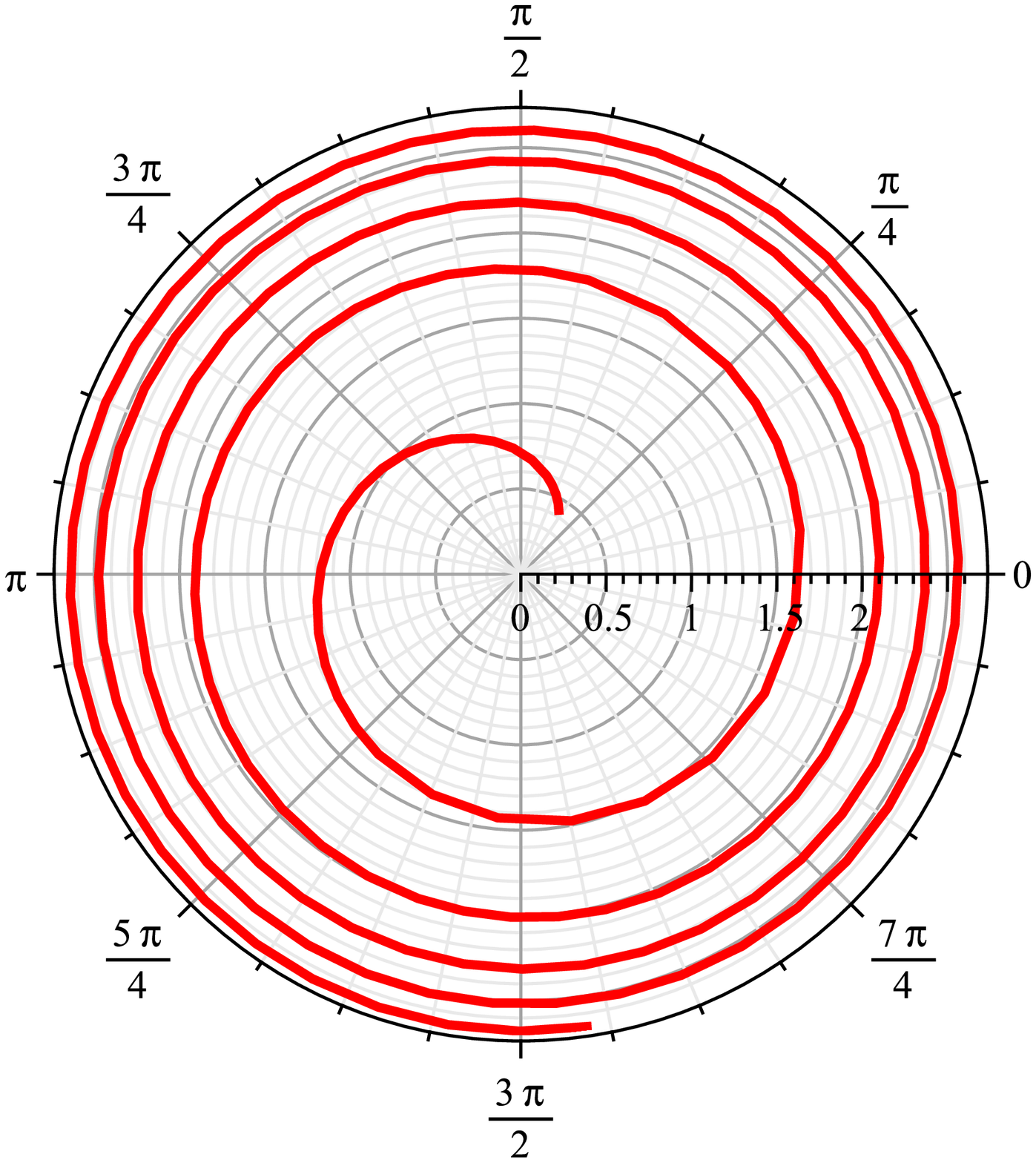} b)
\caption{\small{\textbf{a}) The behavior of coordinate time with respect to initial velocity. \textbf{b}) The spiral behavior of
time, with respect to initial velocity.}}}
\label{t-v}
\end{figure}
We can find out that this behavior, construct a spiral with a
varying interior radius.

Another type of illustrations, is to plot the spatial coordinate
$X$, with respect to coordinate time $t$. To do this, we use the
the general relation (\ref{final-trajectory}), which includes the
effects of the radiation reaction on particle's trajectory. Figure
3 shows the so called behavior.
\begin{figure}[htp]
\center{\includegraphics[width=6cm]{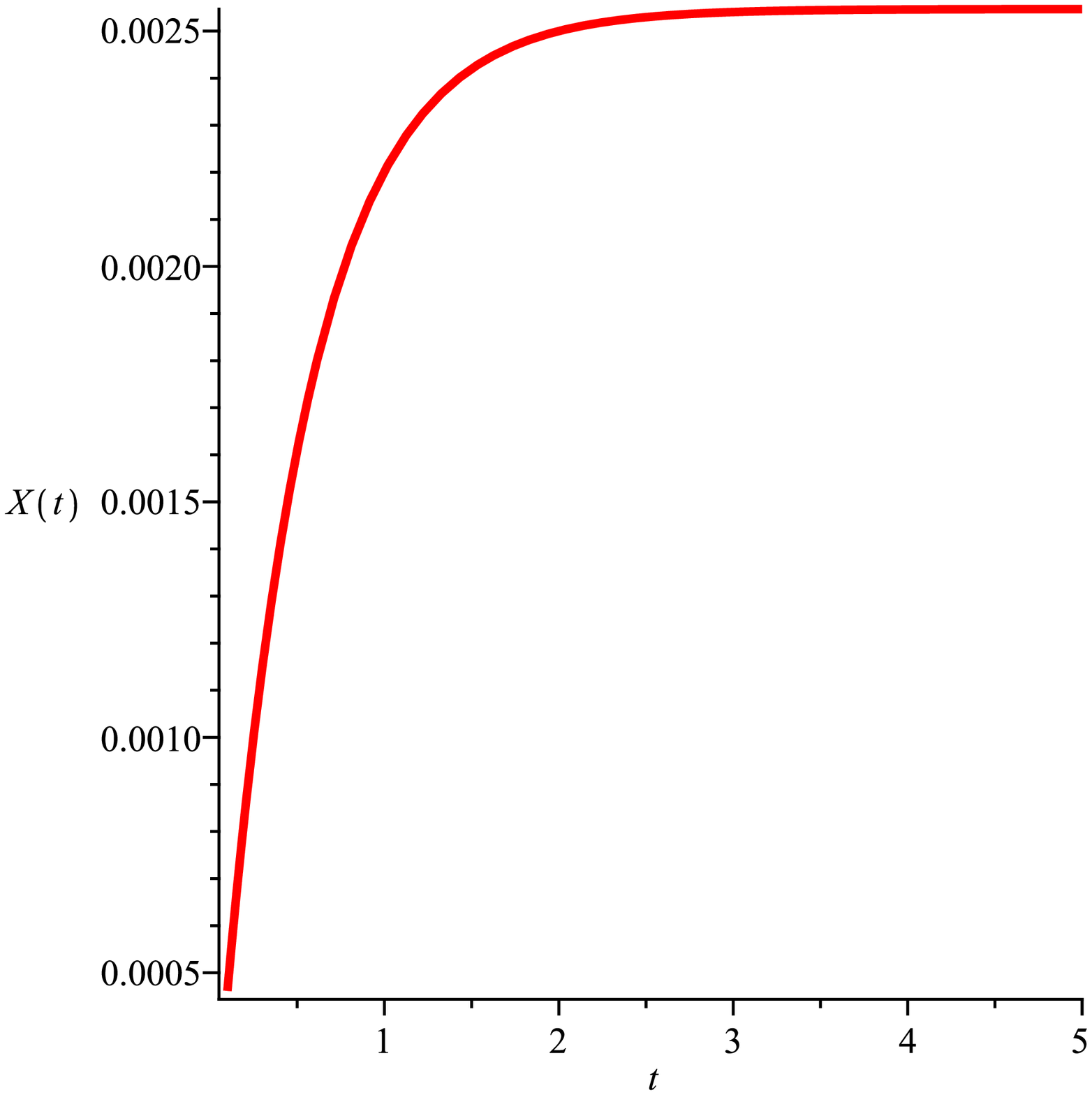} a)
\hfil
\includegraphics[width=6cm]{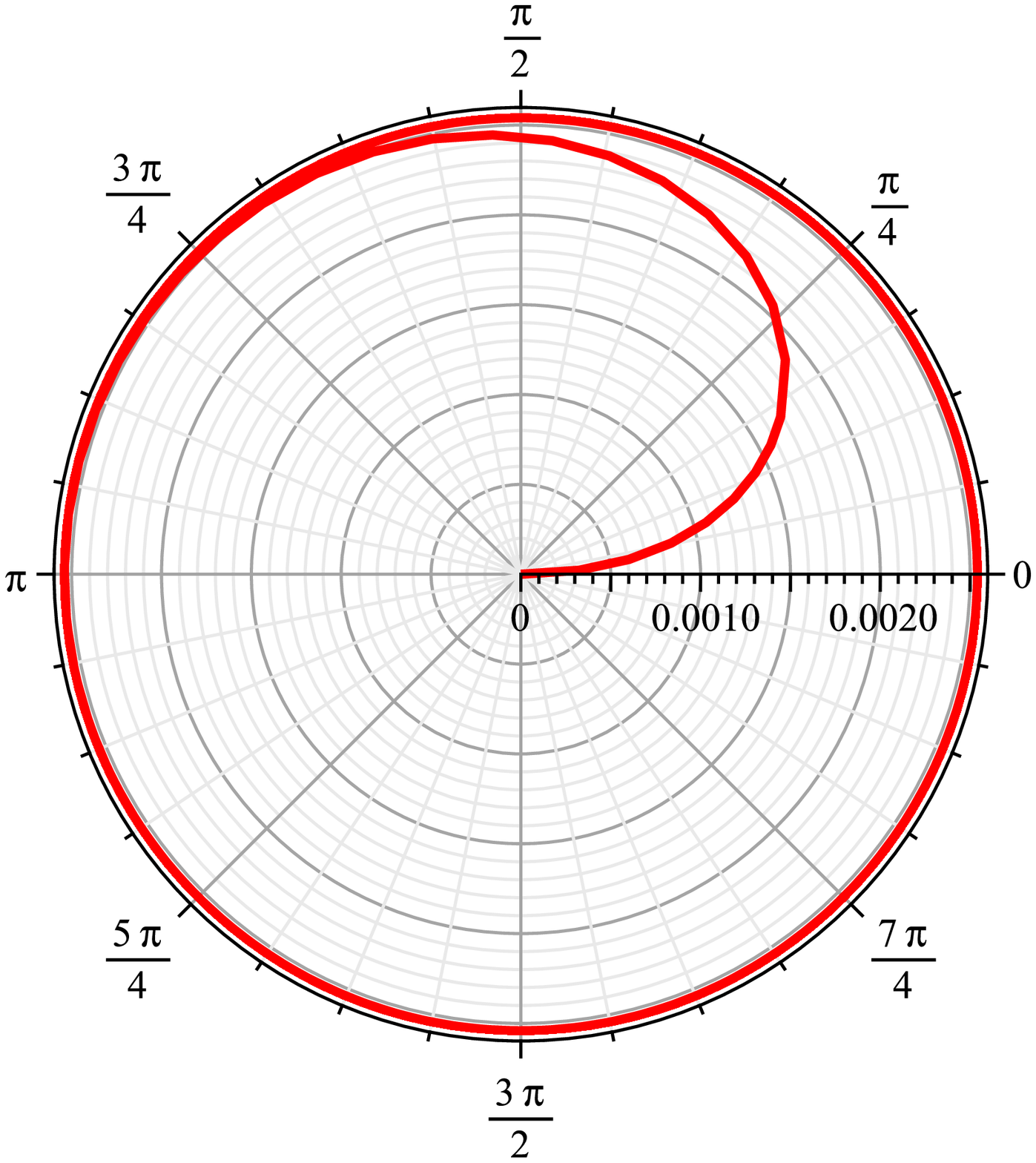} b)
\caption{\small{\textbf{a}) The behavior of spatial coordinates
$X$ with respect to coordinate time $t$. \textbf{b}) The shape of
the test particle's trajectory, in non-relativistic limit.}}}
\label{t-v0} \label{X-t}
\end{figure}
According to Figures 2 and 3, we will see that how the effects of
radiation reaction, can make the trajectories unstable, and also
causes a drop on the canter of potential.

\section{Conclusion}
It seems that the Li\'{e}nard - Wiechert retarded potentials are
not completely sufficient when the trajectories for charged
particles are considered in curved spacetimes \cite{poi2}. In this
article, we firstly reviewed the radiation reaction, caused by
particle's radiation and derived some expressions, relating the
spatial coordinates to coordinate time, in a de Sitter spacetime.
Afterwards, by using a Lagrangian formalism, we derived another
expression for the coordinate time, with respect to the proper
time. Using these relations, we finally plotted the shapes of the
trajectories, for an accelerated charged object, while falling
freely in de Sitter spacetime.\\\\

\textbf{\large{\textbf{\textbf{Acknowledgements}}}} This work was
supported under a research grant by Payame Noor University.


\begin{thebibliography}{0}
\bibitem{Fathi} M. Fathi · M. Tanhayi-Ahari · M.R. Tanhayi · F.
Tavakoli, \emph{Radiation Effects on Geodesics in de Sitter Space:
A Classical Approach}, Int. J. Theor. Phys. (2012) 51:1938-1945.

\bibitem{abraham}M. Abraham and R. Becker, \emph{Theorie der Electrizit\"{at}}, Vol. II, (Springer, Leipzig,
1933).\\ H. A. Lorentz, \emph{Theory of electrons}, (Dover, New
York, 1952).

\bibitem{poisson} F. Rohrlich,
\emph{The dynamics of a charged sphere and the electron.} Am. J.
Phys 65 (11) p. 1051 (1997).

\bibitem{Winberg} Qingming Cheng, \emph{de Sitter space}, in Hazewinkel,
Michiel, Encyclopaedia of Mathematics, Springer (2001), ISBN
978-1556080104.

\bibitem{tanhayi}  M.R. Tanhayi, M. Fathi, M.V. Takook, \emph{Observable quantities in Weyl
gravity}, Mod. Phys. Lett. A, Vol. 26, No. 32 (2011) 2403-2410.

\bibitem{Fathi1} F. Payandeh, M. Fathi, \emph{Spherical Solutions due to the Exterior Geometry
of a Charged Weyl Black Hole}, Int. J. Theor. Phys. (2012)
51:2227-2236.

\bibitem{takook} M. V. Takook, Ph.D. thesis, Universit Paris P.M. Curie
(Paris 6) (1997).

\bibitem{Wheeler} C.W. Misner, K.S. Thorne and J.A. Wheeler, \emph{Gravitation},
Freeman (1973).

\bibitem{f} F. Rohrlich, Am. J. Phys. 68, 12 (2000).


\bibitem{f2} D. V. Gal'tsov, P. Spirin, Grav. Cosmol. 13241-252,
(2007).

\bibitem{landa} L. D. Landau and E. D.
Lifshitz, \emph{The classical Theory Of Fields}, Pergamon Peress,
Fourth English eddition (1975).

\bibitem{cu} Dmitri Gal'tsov, \emph{Radiation reaction and energy-momentum
conservation, "Mass and Motion in General Relativity"}, eds. L.
Blanchet, A. Spallicci and B. Whiting, Springer Series:
Fundamental Theories of Physics, Vol. 162, pp. 367-393, 2011.

\bibitem{curved-po} M.J. Pfenning, E. Poisson, \emph{Scalar, electromagnetic, and
gravitational self-forces in weakly curved spacetimes},
Phys.Rev.D65:084001,2002.

\bibitem{Greiner} Greiner W. Classical Mechanics: Sytem of Particles and Hamiltonian Dynamics. Springer-Verlag, New York, Inc.,
(2003).

\bibitem{poi2} E. Poisson, \emph{Constructing the self-force},
arXiv:0909.2994v1.

\bibitem{lyle} S. N. Lyle, \emph{Self-Force and Inertia: Old Light on New Ideas}, Lect. Notes Phys.
796 (Springer, Berlin Heidelberg 2010), DOI
10.1007/978-3-642-04785-5.

\end{thebibliography}
\end{document}